\documentclass[floatfix,aps,superscriptaddress,prd,amsmath,nofootinbib,preprintnumbers,twocolumn]{revtex4-1}
\usepackage{amssymb}
\usepackage{amsmath}
\usepackage{graphicx}
\usepackage{makecell}
\usepackage[colorlinks=true, citecolor=blue, linkcolor=blue, urlcolor=blue]{hyperref}

\begin{document}

\title{Measuring $H_0$ from low-$z$ datasets}

\author{Xue Zhang}
\affiliation{CAS Key Laboratory of Theoretical Physics, Institute of Theoretical Physics, Chinese Academy of Sciences, Beijing 100190, China}

\author{Qing-Guo Huang}
\affiliation{CAS Key Laboratory of Theoretical Physics, Institute of Theoretical Physics, Chinese Academy of Sciences, Beijing 100190, China}
\affiliation{School of Physical Sciences, University of Chinese Academy of Sciences, No. 19A Yuquan Road, Beijing 100049, China}

\date{\today}

\begin{abstract}
Adopting the low-redshift observational datasets, including the Pantheon sample of Type Ia supernovae, baryon acoustic oscillation measurements, and the tomographic Alcock-Paczynski method, we determine the Hubble constant to be $67.95^{+0.78}_{-1.03}$ , $69.81^{+2.22}_{-2.70}$ and $66.75^{+3.42}_{-4.23}$ km s$^{-1}$ Mpc$^{-1}$ at 68\% confidence level in the $\Lambda$CDM, $w$CDM and $w_0w_a$CDM models, respectively. Compared to the Hubble constant given by Riess et al. in 2019, we conclude that the new physics beyond the standard $\Lambda$CDM model is needed if all of these datasets are reliable.
\end{abstract}

\maketitle

The Hubble constant $H_0$ represents the expansion rate of the universe at present and is closely related to the age of the universe. The accurate measurement of Hubble constant is crucial for modern cosmology. However, different cosmological observations give diverse values of Hubble constant in literature.
Up to now, there are two methods to measure the Hubble constant. One is to directly measure the Hubble constant based on distance ladder estimates of Cepheids and so on. The other is to globally fit the Hubble constant under the assumption of a cosmological model, for example the ``standard" $\Lambda$CDM model. Some results are listed in Table \ref{tab:data}.

\begin{table}[htbp] 
\caption{The measurements of $H_0$ from different datasets}
\centering
\begin{tabular}{ c | c | c | c }
  \hline
  Data & Model & \makecell[c]{$H_0$ \\ (km s$^{-1}$ Mpc$^{-1}$)} & Ref. 
  \\\hline
  HST Key Project \footnote{In 2001 Freedman et al. determined $H_0$ from the final results of the Hubble Space Telescope (HST) Key Project, based on the Cepheid calibration of secondary distance methods.} & - & $72\pm8$ & \cite{Freedman:2000cf} 
  \\
  MCP \footnote{The Megamaser Cosmology Project (MCP) conducted very long baseline interferometry (VLBI) observations
of H$_2$O masers in the accretion disk of the supermassive black holes at the center of the galaxy to measure the geometric distances which gave an angular diameter distance to the galaxy.} & - & $68.9\pm7.1$ & \cite{Reid:2008nm} 
  \\
  CHP \footnote{Using the HST Key Project distance ladder technique with the Spitzer Space Telescope, Freedman et al. reported a significantly improved result of $H_0$ from the Carnegie Hubble Program (CHP).} & - & $74.3\pm2.1$ & \cite{Freedman:2012ny} 
  \\
  LIGO\&Virgo \footnote{Standard-siren measurements from GW170817.} & - & $70.0_{-8.0}^{+12.0}$ & \cite{Abbott:2017xzu}
  \\
  SH0ES \footnote{SH0ES (Supernovae $H_0$ for the Equation of State) team measured $H_0$ based on distance ladder estimates of Cepheids.} & - & $74.03\pm1.42$ & \cite{Riess:2019cxk} 
  \\\hline
  WMAP \footnote{WMAP (Wilkinson Microwave Anisotropy Probe).} & $\Lambda$CDM & $70.0\pm2.2$ & \cite{Hinshaw:2012aka} 
  \\
  BAO+AP \footnote{Cheng et al. proposed that $H_0$ can be determined by combining the low and high-redshift baryonic acoustic oscillation (BAO) distance measurements derived from galaxy surveys in \cite{Cheng:2014kja}, and Alcock-Paczynski (AP) effect tightens the constraints on $\Omega_m$ and hence improves the measurement of $H_0$.} & $\Lambda$CDM & $67.78^{+1.21}_{-1.86}$ & \cite{Zhang:2018jfu}
  \\
  Planck \footnote{Final full-mission Planck satellite measurements of the Cosmic Microwave Background (CMB) anisotropies. } & $\Lambda$CDM & $67.27\pm0.60$ & \cite{Aghanim:2018eyx}
  \\
  WMAP+BAO & $\Lambda$CDM & $68.36^{+0.53}_{-0.52}$ & \cite{Zhang:2018air}
  \\
  H0LiCOW \footnote{The H0LiCOW collaboration ($H_0$ Lenses in COSMOGRAIL's Wellspring) presented a new method based on the measurements of time delays for the gravitational lens systems.} & $\Lambda$CDM & $73.3^{+1.7}_{-1.8}$ & \cite{Wong:2019kwg} 
\\\hline
\end{tabular}
\label{tab:data}
\end{table}

From Table \ref{tab:data}, we see that there is an obvious tension between the local measurement on the Hubble constant from SH0ES and the results from globally fitting the cosmic microwave background (CMB), including Wilkinson Microwave Anisotropy Probe (WMAP) and Planck, and baryonic acoustic oscillation (BAO). In particular, the tension between SH0ES \cite{Riess:2019cxk} and Planck \cite{Aghanim:2018eyx} has been at around $4.4 \sigma$ level. And both the combinations of WMAP + BAO datasets \cite{Zhang:2018air} and BAO + Alcock-Paczynski (AP) datasets \cite{Zhang:2018jfu} give similar Hubble constant with Planck in the standard $\Lambda$CDM cosmology. On the other hand, recently a new measurement of the Hubble constant, namely H0LiCOW \cite{Wong:2019kwg}, in the $\Lambda$CDM model from a joint analysis of six gravitationally lensed quasars with measured time delays which is completely independent of both supernovae and CMB analyses roughly recovers the Hubble constant from SH0ES.

 Even though the tension on the Hubble constant from diverse measurements might arise from the uncertain systematic errors in the datasets, it is still important to propose new methods to measure the Hubble constant. In this letter we suggest to determine the Hubble constant by adopting the low-redshift datasets, including BAO, AP and supernovae data, in the models beyond the standard $\Lambda$CDM model which are supposed to relax the cosmological model dependence.

 Here we only adopt the low-redshift datasets, including BAO measurements from 6dFGS, MGS, DR12, DR14 and eBOSS DR14 Ly$\alpha$, Pantheon sample of Type Ia supernovae (SNe Ia), and AP effect from the BOSS DR12 galaxies.
 \begin{itemize}
  \item The Pantheon sample, as the largest confirmed SNe Ia sample, which includes 1048 SNe Ia and covers the redshift range of $0.01<z<2.3$ \cite{Scolnic:2017caz}.
  \item The BAO measurements from the 6dF Galaxy Survey (6dFGS) at $z_\mathrm{eff}=0.106$ \cite{Beutler:2011hx},
        the SDSS DR7 main Galaxy sample (MGS) at $z_\mathrm{eff}=0.15$ \cite{Ross:2014qpa},
        the BOSS DR12 sample \cite{Wang:2016wjr} at $z_\mathrm{eff} = 0.31, 0.36, 0.40, 0.44, 0.48, 0.52, 0.56, 0.59, 0.64$.
        the eBOSS DR14 quasar sample at $z_\mathrm{eff} = 1.52$ \cite{Ata:2017dya},
        the eBOSS DR14 Ly$\alpha$ at $z_\mathrm{eff} = 2.34$ \cite{Agathe:2019vsu}.
  \item The tomographic AP method to the BOSS DR12 galaxies ($0.15<z<0.693$) \cite{Li:2016wbl}.
\end{itemize}

Since the late-time dynamics of the Universe is dominated by dark energy, the properties of dark energy significantly affect the expansion history of the Universe. Therefore, we constrain the Hubble constant in the standard $\Lambda$CDM model and then extend to $w$CDM and $w_0w_a$CDM models. We use the CosmoMC software \cite{Lewis:2002ah} to obtain the Markov Chain Monte Carlo (MCMC) samples. Here we assume a fiducial cosmology, namely $\Omega_\mathrm{b}h^2=0.02236$ \cite{Aghanim:2018eyx}. The changes of $\Omega_\mathrm{b}h^2$ within its error bars do not substantially shift our results.

\begin{figure}[htbp]
\centering
\includegraphics[width=.45\textwidth]{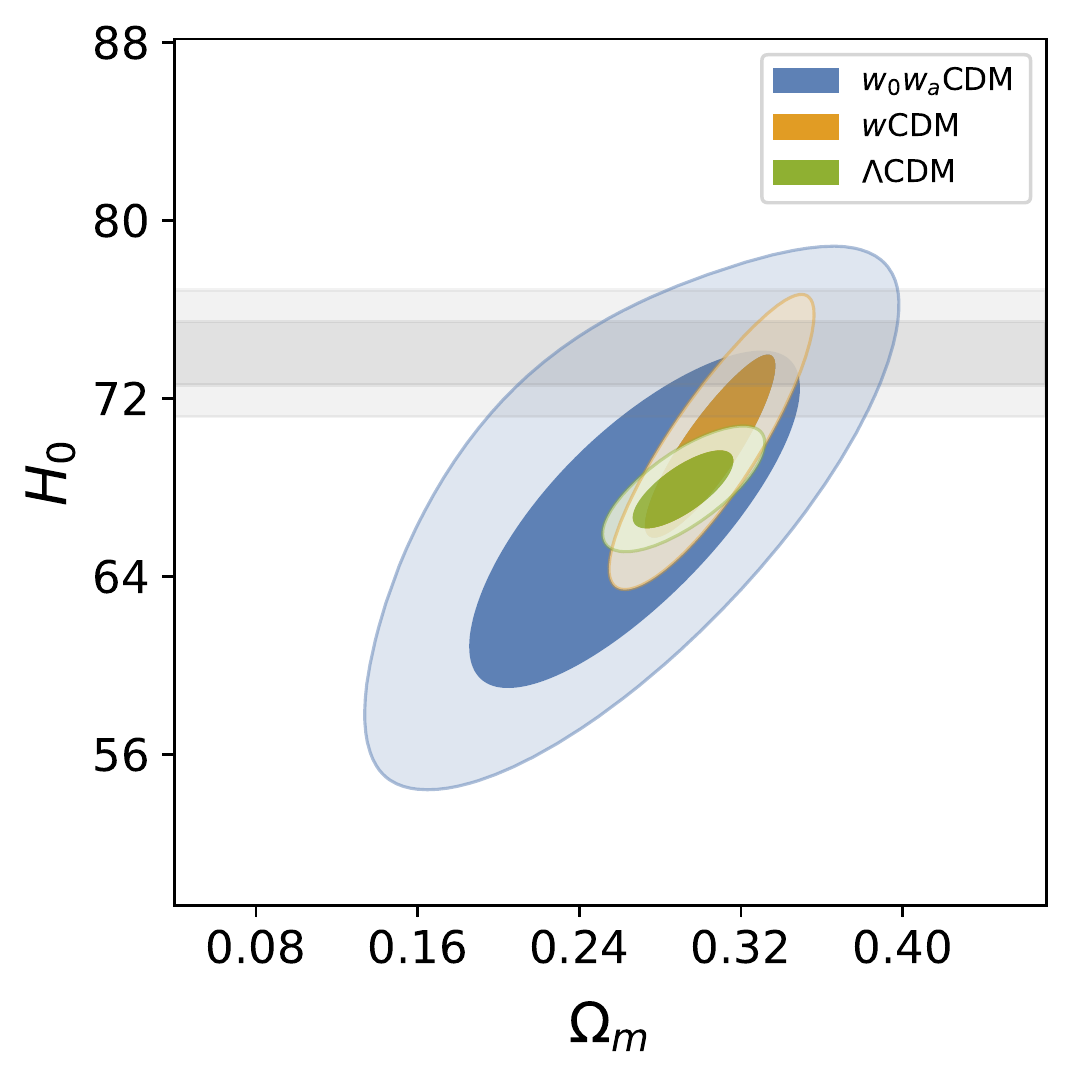}\\
\includegraphics[width=.45\textwidth]{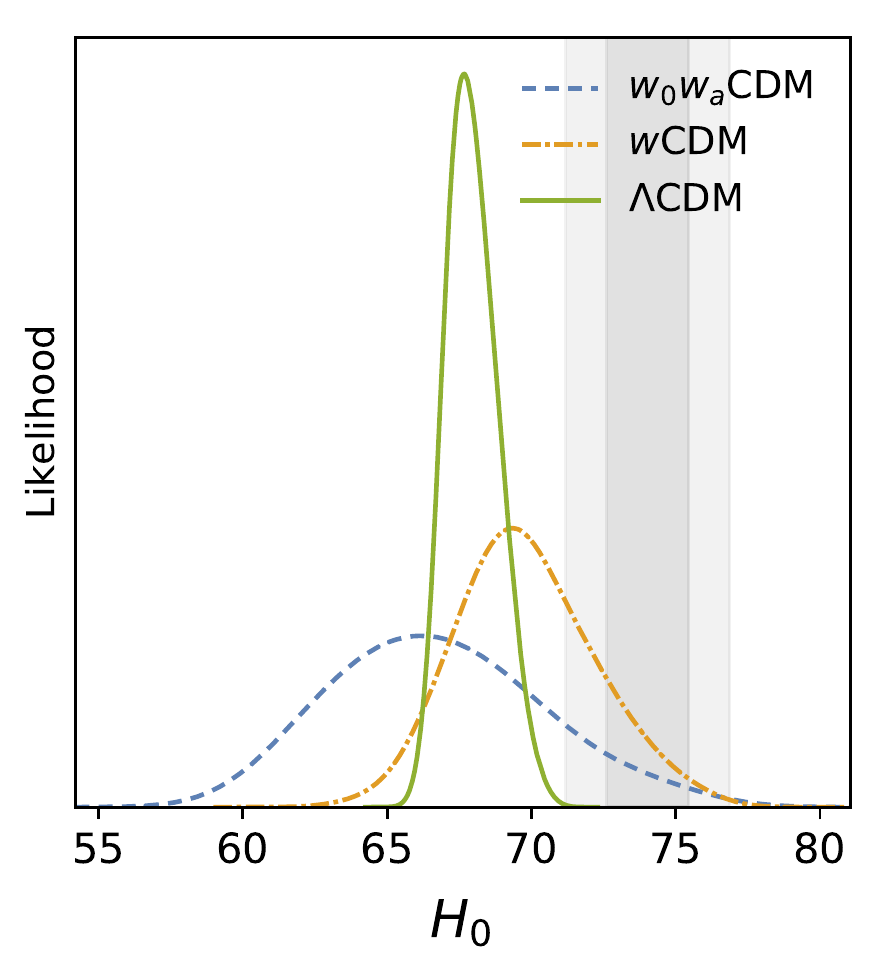}
\caption{Top panel: the marginalized contour (68\% CL and 95\% CL) of parameters $H_0$ and $\Omega_\mathrm{m}$
for the $\Lambda$CDM, $w$CDM and $w_0w_a$CDM models from Pantheon+BAO+AP datasets.
Down panel: the likelihood distributions of Hubble constant $H_0$
for the $\Lambda$CDM, $w$CDM and $w_0w_a$CDM models from Pantheon+BAO+AP datasets.
The gray bands represent the constraint from SH0ES in ref. \cite{Riess:2019cxk}.}
\label{fig:H0}
\end{figure}

We perform a likelihood analysis to place constraints on the parameters space in the $\Lambda$CDM, $w$CDM and $w_0w_a$CDM models from Pantheon+BAO+AP datasets. The total likelihood $\chi^2$ for the datasets can be constructed by
$\chi^2_{\mathrm{total}}=\chi^2_{\mathrm{Pantheon}}+\chi^2_{\mathrm{BAO}}+\chi^2_{\mathrm{AP}}$. Our results are illustrated in Figure \ref{fig:H0} and Table \ref{tab:result}.

\begin{table}[htbp]
\caption{Parameter 68\% intervals and $\chi^2$ for the $\Lambda$CDM, $w$CDM and $w_0w_a$CDM models from Pantheon+BAO+AP datasets}
\centering
\begin{tabular}{l c c c }
  \hline
  Parameter & $\Lambda$CDM & $w$CDM & $w_0w_a$CDM
  \\\hline
  $w_0$
  & -1
  & $-1.05\pm0.06$
  & $-1.05\pm0.07$
  \\
  $w_a$
  & -
  & -
  & $0.59^{+0.52}_{-0.24}$
  \\\hline
  $\Omega_\mathrm{m}$
  & $0.292^{+0.011}_{-0.015}$
  & $0.305^{+0.019}_{-0.021}$
  & $0.265^{+0.045}_{-0.035}$
  \\
  $H_0$
  & $67.95^{+0.78}_{-1.03}$
  & $69.81^{+2.22}_{-2.70}$
  & $66.75^{+3.42}_{-4.23}$
  \\\hline
  $\chi^2_{\mathrm{Pantheon}}$
  & 1036.42
  & 1036.68
  & 1036.77
  \\
  $\chi^2_{\mathrm{BAO}}$
  & 17.34
  & 18.19
  & 17.10
  \\
  $\chi^2_{\mathrm{AP}}$
  & 73.41
  & 72.91
  & 72.16
  \\
  $\chi^2_{\mathrm{total}}$
  & 1127.17
  & 1127.79
  & 1126.03
\\\hline
\end{tabular}
\label{tab:result}
\end{table}

The Hubble constant can be precisely determined by the low-redshift observational data in the $\Lambda$CDM model. Here we present an $1.3\%$ precision measurement:
\begin{equation}
H_0=67.95^{+0.78}_{-1.03},
\end{equation}
which has a tension with the Hubble constant given by Riess et al. (2019) at the $3.75\sigma$ level. However, since the measurement of the Hubble constant using Pantheon+BAO+AP datasets depends on the cosmological model, we extend to $w$CDM and $w_0w_a$CDM models, and find
\begin{equation}
H_0=69.81^{+2.22}_{-2.70}~(3.5 \%~\text{precision}).
\end{equation}
for the $w$CDM model, and
\begin{equation}
H_0=66.75^{+3.42}_{-4.23}~(5.7 \%~\text{precision})
\end{equation}
for the $w_0w_a$CDM model. The measurements have worse precision for more general dark energy models, but can alleviate the tension with $H_0$ by Riess et al. (2019).

To summarize, we combine the low-redshift observational datasets from the Pantheon sample of SNe Ia sample,
the BAO measurements (including the 6dFGS, the SDSS DR7 MGS the BOSS DR12, the eBOSS DR14 quasar, the eBOSS DR14 Ly$\alpha$), and the tomographic Alcock-Paczynski method to the BOSS DR12 galaxies, and globally determine the Hubble constant in three cosmological models.
For the standard $\Lambda$CDM model, the Hubble constant is $H_0=67.95^{+0.78}_{-1.03}$ km s$^{-1}$ Mpc$^{-1}$ which is nicely consistent the measurement from CMB data, but has a $3.75\sigma$ tension with that given by Riess et al. (2019). Such a tension can be significantly relaxed in the $w$CDM model and $w_0w_a$CDM model. In fact, a dynamical dark energy model can also relax the tension between the local measurement on the Hubble constant and the globally fitting the CMB and BAO data \cite{Qing-Guo:2016ykt}.

Even though the true value of the Hubble constant is still unknown, we conclude that the tension between the local measurement and the globally fitting from CMB and other low-redshift cosmological data strongly implies that we need new physics beyond the standard $\Lambda$CDM cosmology. It may imply for a dynamical dark energy \cite{Qing-Guo:2016ykt,dynamicalDE}, or other explanations \cite{others}. In a word, more study on the Hubble constant is needed in the future.

Acknowledgments. We acknowledge the use of HPC Cluster of ITP-CAS.
This work is supported by grants from NSFC
(grant No. 11690021, 11975019, 11847612),
the Strategic Priority Research Program of Chinese Academy of Sciences
(Grant No. XDB23000000, XDA15020701), and Key Research Program of Frontier Sciences, CAS, Grant NO. ZDBS-LY-7009.


\begin{thebibliography}{99}

\bibitem{Freedman:2000cf}
  W.~L.~Freedman et al. (HST Collaboration),
  Astrophys.\ J.\  {\bf 553}, 47 (2001),
  astro-ph/0012376.


\bibitem{Reid:2008nm}
  M.~J.~Reid, J.~A.~Braatz, J.~J.~Condon, L.~J.~Greenhill, C.~Henkel, and K.~Y.~Lo,
  Astrophys.\ J.\  {\bf 695}, 287 (2009),
  arXiv:0811.4345.

\bibitem{Freedman:2012ny}
  W.~L.~Freedman, B.~F.~Madore, V.~Scowcroft, C.~Burns, A.~Monson, S.~E.~Persson, M.~Seibert, and J.~Rigby,
  Astrophys.\ J.\  {\bf 758}, 24 (2012),
  arXiv:1208.3281.

\bibitem{Abbott:2017xzu}
  B.~P.~Abbott et al. (LIGO Scientific and Virgo and 1M2H and Dark Energy Camera GW-E and DES and DLT40 and Las Cumbres Observatory and VINROUGE and MASTER Collaborations),
  Nature {\bf 551}, 85 (2017),
  arXiv:1710.05835.

\bibitem{Riess:2019cxk}
  A.~G.~Riess, S.~Casertano, W.~Yuan, L.~M.~Macri, and D.~Scolnic,
  Astrophys.\ J.\  {\bf 876}, 85 (2019),
  arXiv:1903.07603.

\bibitem{Hinshaw:2012aka}
  G.~Hinshaw et al. (WMAP Collaboration),
  Astrophys.\ J.\ Suppl.\  {\bf 208}, 19 (2013),
  arXiv:1212.5226.


\bibitem{Zhang:2018jfu}
  X.~Zhang, Q.~G.~Huang, and X.~D.~Li,
  Mon.\ Not.\ R.\ Astron.\ Soc.\  {\bf 483}, 1655 (2019),
  arXiv:1801.07403.

\bibitem{Cheng:2014kja}
  C.~Cheng, and Q.~G.~Huang,
  Sci.\ China-Phys.\ Mech.\ Astron.\  {\bf 58}, 599801 (2015),
  arXiv:1409.6119.


\bibitem{Aghanim:2018eyx}
  N.~Aghanim et al. (Planck Collaboration),
  arXiv:1807.06209.

\bibitem{Zhang:2018air}
  X.~Zhang, and Q.~G.~Huang,
  Commun.\ Theor.\ Phys.\  {\bf 71}, 826 (2019),
  arXiv:1812.01877.

\bibitem{Wong:2019kwg}
  K. C. Wong, S. H. Suyu, G. C. F. Chen, C. E. Rusu, M. Millon, D. Sluse, V. Bonvin, C. D. Fassnacht, S. Taubenberger, M. W. Auger,
  S. Birrer, J. H. H. Chan, F. Courbin, S. Hilbert, O. Tihhonova, T. Treu, A. Agnello, X. Ding, I. Jee, E. Komatsu,
  A. J. Shajib, A. Sonnenfeld, R. D. Blandford, L. V. E. Koopmans, P. J. Marshall, and G. Meylan,
  arXiv:1907.04869.

\bibitem{Scolnic:2017caz}
  D. M. Scolnic, D. O. Jones, A. Rest, Y. C. Pan, R. Chornock, R. J. Foley, M. E. Huber, R. Kessler, G. Narayan, A. G. Riess,
  S. Rodney, E. Berger, D. J. Brout, P. J. Challis, M. Drout, D. Finkbeiner, R. Lunnan, R. P. Kirshner, N. E. Sanders, E. Schlafly,
  S. Smartt, C. W. Stubbs, J. Tonry, W. M. Wood-Vasey, M. Foley, J. Hand, E. Johnson, W. S. Burgett, K. C. Chambers, P. W. Draper,
  K. W. Hodapp, N. Kaiser, R. P. Kudritzki, E. A. Magnier, N. Metcalfe, F. Bresolin, E. Gall, R. Kotak, M. McCrum, and K. W. Smith,
  Astrophys.\ J.\  {\bf 859}, 101 (2018),
  arXiv:1710.00845.

\bibitem{Beutler:2011hx}
  F. Beutler, C. Blake, M. Colless, D.H. Jones, L.S. Smith, L. Campbell, Q. Parker, W. Saunders, and F. Watson
  Mon.\ Not.\ R.\ Astron.\ Soc.\  {\bf 416}, 3017 (2011),
  arXiv:1106.3366.
\bibitem{Ross:2014qpa}
  A.~J.~Ross, L.~Samushia, C.~Howlett, W.~J.~Percival, A.~Burden, and M.~Manera,
  Mon.\ Not.\ R.\ Astron.\ Soc.\  {\bf 449}, 835 (2015),
  arXiv:1409.3242.
\bibitem{Wang:2016wjr}
  Y.~Wang et al. (BOSS Collaboration),
  Mon.\ Not.\ R.\ Astron.\ Soc.\  {\bf 469}, 3762 (2017),
  arXiv:1607.03154.
\bibitem{Ata:2017dya}
  M. Ata, F. Baumgarten, J. Bautista, F. Beutler, D. Bizyaev, M. R. Blanton, J. A. Blazek, A. S. Bolton, J. Brinkmann, J. R. Brownstein,
  E. Burtin, C. H. Chuang, J. Comparat, K. S. Dawson, A. de la Macorra, W. Du,  H. du Mas des Bourboux, D. J. Eisenstein, H. Gil-Mar\'{\i}n, K. Grabowski,
  J. Guy, N. Hand, S. Ho, T. A. Hutchinson, M. M. Ivanov, F. S. Kitaura, J. P. Kneib, P. Laurent, J. M. Le Goff, J. E. McEwen,
  E. M. Mueller, A. D. Myers, J. A. Newman, N. Palanque-Delabrouille, K. Pan, I. Paris, M. Pellejero-Ibanez, W. J. Percival, P. Petitjean, F. Prada,
  A. Prakash, S. A. Rodr\'{\i}guez-Torres, A. J. Ross, G. Rossi, R. Ruggeri, A. G. Sanchez, S. Satpathy, D. J. Schlegel, D. P. Schneider, H. J. Seo,
  A. Slosar, A. Streblyanska, J. L. Tinker, R. Tojeiro, M. V. Magana, M. Vivek, Y. Wang, C. Yeche, L. Yu, and P. Zarrouk,
  C. Zhao, G. B. Zhao, and F. Zhu,
  Mon.\ Not.\ R.\ Astron.\ Soc.\  {\bf 473}, 4773 (2018),
  arXiv:1705.06373.
\bibitem{Agathe:2019vsu}
  V. de Sainte Agathe, C. Balland, H. du Mas des Bourboux, N. G. Busca, M. Blomqvist, J. Guy, J. Rich, A. Font-Ribera, M. M. Pieri, J. E. Bautista,
  K. Dawson, J. M. Le Goff, A. de la Macorra, N. Palanque-Delabrouille, W. J. Percival, I. P\'{e}rez-R\`{a}fols, D. P. Schneider, A. Slosar, and C. Y\`{e}che,
  Astron.\ Astrophys.\  {\bf 629}, A85 (2019),
  arXiv:1904.03400.

\bibitem{Li:2016wbl}
  X.~D.~Li, C.~Park, C.~G.~Sabiu, H.~Park, D.~H.~Weinberg, D.~P.~Schneider, J.~Kim, and S.~E.~Hong,
  Astrophys.\ J.\  {\bf 832}, 103 (2016),
  arXiv:1609.05476.

\bibitem{Lewis:2002ah}
  A.~Lewis, and S.~Bridle,
  Phys.\ Rev.\ D {\bf 66}, 103511 (2002),
  astro-ph/0205436.

\bibitem{Qing-Guo:2016ykt}
  Q.~G.~Huang, and K.~Wang,
  Eur.\ Phys.\ J.\ C {\bf 76}, 506 (2016),
  arXiv:1606.05965.


\bibitem{dynamicalDE}
  E.~Di Valentino, A.~Melchiorri, E.~V.~Linder, and J.~Silk,
  Phys.\ Rev.\ D {\bf 96}, 023523 (2017),
  arXiv:1704.00762;
  J.~Ryan, Y.~Chen, and B.~Ratra,
  Mon.\ Not.\ R.\ Astron.\ Soc.\  {\bf 488}, 3844 (2019),
  arXiv:1902.03196;
  Z. Zhang, G. Gu, X. Wang, Y. H. Li, C. G. Sabiu, H. Park, H. Miao, X. Luo, F. Fang, and X. D. Li,
  Astrophys.\ J.\  {\bf 878}, 137 (2019),
  arXiv:1902.09794;
  C.~G.~Park, and B.~Ratra,
  arXiv:1908.08477.






\bibitem{others}
  S.~Kumar, and R.~C.~Nunes,
  Phys.\ Rev.\ D {\bf 94}, 123511 (2016),
  arXiv:1608.02454;
  S.~Vagnozzi, E.~Giusarma, O.~Mena, K.~Freese, M.~Gerbino, S.~Ho, and M.~Lattanzi,
  Phys.\ Rev.\ D {\bf 96}, 123503 (2017),
  arXiv:1701.08172;
  X.~Zhang,
  Sci.\ China-Phys.\ Mech.\ Astron.\  {\bf 60}, 060421 (2017),
  arXiv:1702.05010;
  E.~Di Valentino, A.~Melchiorri, and O.~Mena,
  Phys.\ Rev.\ D {\bf 96}, 043503 (2017),
  arXiv:1704.08342;
  S.~M.~Feeney, H.~V.~Peiris, A.~R.~Williamson, S.~M.~Nissanke, D.~J.~Mortlock, J.~Alsing, and D.~Scolnic,
  Phys.\ Rev.\ Lett.\  {\bf 122}, 061105 (2019),
  arXiv:1802.03404;
  H.~Miao, and Z.~Huang,
  Astrophys.\ J.\  {\bf 868}, 20 (2018),
  arXiv:1803.07320;
  M.~Ishak,
  Living Rev.\ Rel.\  {\bf 22}, 1 (2019),
  arXiv:1806.10122;
  E.~\'{O} Colg\'{a}in, M.~H.~P.~M.~van Putten, and H.~Yavartanoo,
  Phys.\ Lett.\ B {\bf 793}, 126 (2019),
  arXiv:1807.07451;
  R.~Y.~Guo, J.~F.~Zhang, and X.~Zhang,
  J. Cosmol. Astropart. Phys. {\bf 1902}, 054 (2019),
  arXiv:1809.02340.
  H.~Xu, Z.~Huang, Z.~Liu, and H.~Miao,
  Astrophys.\ J.\  {\bf 877},  107 (2019),
  arXiv:1812.09100;
  X.~D.~Li, H.~Miao, X.~Wang, X.~Zhang, F.~Fang, X.~Luo, Q.~G.~Huang, and M.~Li,
  Astrophys.\ J.\  {\bf 875}, 92 (2019),
  arXiv:1903.04757;
  W.~Yang, O.~Mena, S.~Pan, and E.~Di Valentino,
  Phys.\ Rev.\ D {\bf 100}, 083509 (2019),
  arXiv:1906.11697;
  S.~Pan, W.~Yang, E.~Di Valentino, E.~N.~Saridakis, and S.~Chakraborty,
  arXiv:1907.07540;
  S.~Pan, W.~Yang, E.~Di Valentino, A.~Shafieloo, and S.~Chakraborty,
  arXiv:1907.12551;
  S.~Ghosh, R.~Khatri, and T.~S.~Roy,
  arXiv:1908.09843;
  N.~Khadka, and B.~Ratra,
  arXiv:1909.01400;
  M.~Escudero, and S.~J.~Witte,
  arXiv:1909.04044;
  C.~Nicolaou, O.~Lahav, P.~Lemos, W.~Hartley and J.~Braden,
  arXiv:1909.09609;
  E.~Di Valentino, A.~Melchiorri, O.~Mena, and S.~Vagnozzi,
  arXiv:1910.09853.
  W.~Lin, and M.~Ishak,
  arXiv:1909.10991;








\end{thebibliography}
\end{document}